\documentclass[%
reprint,
superscriptaddress,
amsmath,amssymb,
aps,
footinbib,
floatfix
]{revtex4-2}

\usepackage{graphicx}
\usepackage{siunitx}
\usepackage{dcolumn}
\usepackage{bm}
\usepackage{kantlipsum}
\usepackage{hyperref}
\usepackage[mathlines]{lineno}
\usepackage{cleveref}
\usepackage{physics}
\usepackage{subfiles}
\usepackage{xcolor}

\usepackage{bigints}

\DeclareSIUnit\sample{Sa}

\usepackage[sectionbib]{bibunits}
\defaultbibliographystyle{ieeetr}
\defaultbibliography{Ref_PhaseEstimation}

\usepackage{chngcntr}
\begin{document}

\preprint{AIP/123-QED}

\title{Variational quantum algorithm for enhanced continuous variable optical phase sensing}

\newcommand{\addrbigQ}{Center for Macroscopic Quantum States bigQ, Department of Physics, Technical University of Denmark, Fysikvej 307, DK-2800 Kgs.\ Lyngby, Denmark}

\author{Jens A. H. Nielsen}
\affiliation{\addrbigQ}
\author{Mateusz Kicinski}
\affiliation{\addrbigQ}
\author{Tummas N. Arge}
\affiliation{\addrbigQ}
\author{Kannan Vijayadharan}
\affiliation{Department of Information Engineering, University of Padova}
\author{Jonathan Foldager}
\affiliation{Department of Applied Mathematics and Computer Science, Technical University of Denmark}
\author{Johannes Borregaard}
\email{borregaard@fas.harvard.edu}
\affiliation{Qutech, Delft University of Technology}
\affiliation{Department of Physics, Harvard University}
\author{Johannes Jakob Meyer}
\affiliation{Dahlem Center for Complex Quantum Systems, Freie Universität Berlin}
\author{Jonas S. Neergaard-Nielsen}
\affiliation{\addrbigQ}
\author{Tobias Gehring}
\affiliation{\addrbigQ}
\author{Ulrik L. Andersen}
\email{ulrik.andersen@fysik.dtu.dk}
\affiliation{\addrbigQ}

\begin{abstract}
Variational quantum algorithms (VQAs) are hybrid quantum-classical approaches used for tackling a wide range of problems on noisy intermediate-scale quantum (NISQ) devices.
Testing these algorithms on relevant hardware is crucial to investigate the effect of noise and imperfections and to assess their practical value. 
Here, we implement a variational algorithm designed for optimized parameter estimation on a continuous variable platform based on squeezed light, a key component for high-precision optical phase estimation. 
We investigate the ability of the algorithm to identify the optimal metrology process, including the optimization of the probe state and measurement strategy for small-angle optical phase sensing. Two different optimization strategies are employed, the first being a gradient descent 
optimizer using Gaussian parameter shift rules to estimate the gradient of the cost function directly from the measurements. The second strategy involves a gradient-free Bayesian optimizer, fine-tuning the system using the same cost function and trained on the data acquired through the gradient-dependent algorithm. We find that both algorithms can steer the experiment towards the optimal metrology process. However, they find minima not predicted by our theoretical model, demonstrating the strength of variational algorithms in modelling complex noise environments, a non-trivial task.
\end{abstract}

\maketitle

\begin{figure*}
    \centering
    \includegraphics[width = \linewidth]{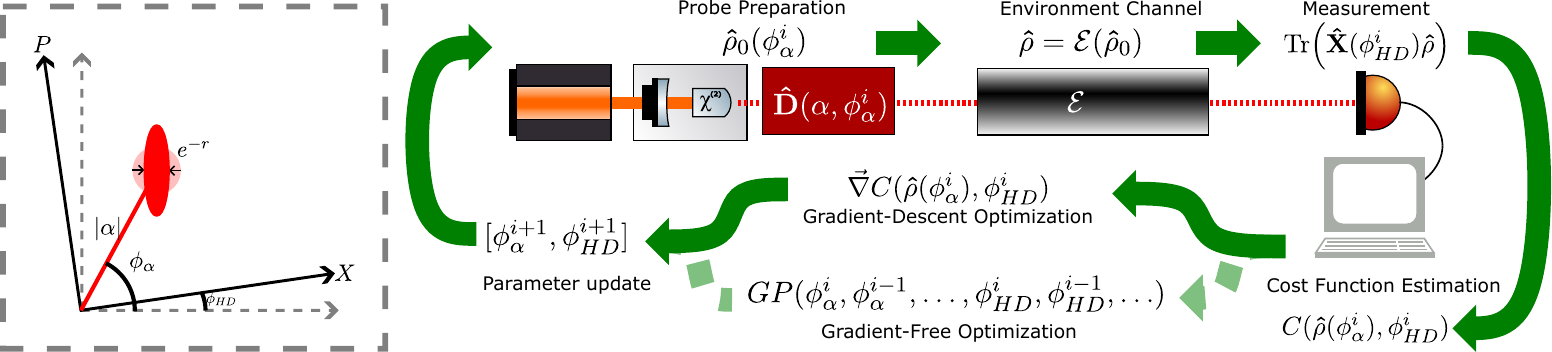}
    \caption{The overall principle of our variational experiment. A probe state is prepared as a squeezed, displaced state. The free parameters of our system are the measurement basis angle, $\phi_{HD}$, and the displacement angle, $\phi_\alpha$, both relative to the squeezing angle $\phi_{r}$. After interacting with the environment, the state is detected in a homodyne detector. The cost function is estimated by varying the measurement basis of the detector. Subsequently, the experiment either estimates the gradient of the cost function to determine the next set of initial parameters or employs a Gaussian Process (GP) in a Bayesian gradient-free optimization algorithm for parameter selection.}
    \label{fig:principle}
\end{figure*}

Hybrid quantum-classical algorithms~\cite{cerezo2021variational,bharti2022noisy} like the quantum approximate optimization algorithm (QAOA) and the variational quantum eigensolver (VQE) show promise for implementation on early quantum devices that are yet incapable of running quantum error correction. There have been several experimental demonstrations of these algorithms on quantum hardware such as trapped ions~\cite{Meth2022,Hempel2018}, neutral atoms~\cite{Graham2022}, and superconducting qubits~\cite{Malley2016,self2021,Zhang2022}. Although these algorithms are suitable for a wide range of optimization problems ranging from electronic structure calculations~\cite{cao2019quantum} to general combinatorial optimization problems~\cite{farhi2014quantum}, the demonstration of a practical quantum advantage compared to purely classical techniques remains an outstanding challenge. 

Quantum metrology, which leverages inherently quantum effects such as entanglement to surpass classical limits on resolution and sensitivity~\cite{Giovannetti2006}, is a promising area for near-term applications of noisy quantum hardware. Numerous theoretical studies have shown that variational methods can be applied to identify optimal non-classical probe states and measurements in the presence of unknown noise processes and imperfections~\cite{kaubruegger2019variational,koczor2020variational-state,yang2020probe,meyer2021variational}. Recent experimental demonstrations with photonic platforms have also highlighted the feasibility of these techniques for multi-parameter sensing in the few photon limit~\cite{Cimini2023}.   

We extend previous investigations to the practically relevant regime of optical phase estimation using a squeezed coherent state and homodyne measurements~\cite{Lawrie2019}. This combination has been successfully deployed to enhance gravitational wave-detection~\cite{LIGO2013,Grote2013}, magnetic field sensing~\cite{Li2018}, and biological imaging~\cite{Taylor2013} beyond classical limits. We demonstrate that a variational algorithm can effectively steer the experiment towards the optimal probe state and measurement basis in the presence of various imperfections like phase fluctuations and loss. 
Our approach involves optimizing the classical Fisher information, a key determinant of metrological precision.
We develop and implement parameter shift rules \cite{Schuld2019} to calculate and differentiate this quantity in continuous variable systems. This allows us to implement two approaches: one that estimates the gradient of the classical Fisher information through additional measurements for gradient descent optimization, and another that employs a gradient-free Bayesian optimizer to further fine-tune the optimization process.\\
Our results confirm the potential of variational techniques for quantum sensing tasks and highlight the remaining challenges for their practical application.
We find that the additional time overhead incurred by calculating the gradient is balanced by the gradient-based optimizer's better handling of slow drifts in the sensing apparatus. Conversely, the gradient-free optimization method requires more fine-tuning of the exploration/exploitation trade-off, but enables faster convergence. Further investigation of variational methods in more complex setups with intricate parametrizations, and a detailed study of the potentials and limitations of different optimization techniques, will be crucial for the widespread adoption of variational metrology schemes.

Our task is to optimize a metrological protocol, which includes preparing a probe state and a suitable measurement, to optimize a small phase imprinted on a mode of a continuous-variable quantum system.
This is of fundamental interest in applied quantum sensing, as continuous-variable quantum systems are widely used and have proven their usefulness in practically relevant scenarios~\cite{LIGO2013,Grote2013,Taylor2013}.

The conventional approach to developing metrological protocols typically starts with a theoretical description, followed by protocol development and implementation. This strategy has multiple downsides: 
A completely faithful model of a system is often hard to devise and when trying to push the system to its limits, every inaccuracy in the theoretical model can negatively impact performance. 
Furthermore, even with a faithful model, the system often depends on parameters that are not stable over time, causing parameter drift. 
Therefore, methods that adapt to the actual physical conditions during execution, without requiring a complete theoretical model, are highly desirable.

Variational quantum algorithms for quantum metrology~\cite{kaubruegger2019variational,koczor2020variational-state,yang2020probe,meyer2021variational} have been proposed for this purpose. In these approaches, a cost function is defined to represent the performance of a metrological protocol and is then optimized. This process does not require a full theoretical model, allowing the protocol to implicitly account unmodeled effects.

Our experiment broadly consists of three stages. First, a displaced squeezed state is prepared as a probe state, where the displacement phase angle relative to the squeezed quadrature, $\phi_{\alpha}$, is a free parameter. We fix the squeezing level $r$ and displacement amplitude $|\alpha|$ to work with probes of a fixed photon number. Next, the state undergoes a phase shift that encodes the parameter to be sensed. Finally, we perform homodyne detection, where the homodyne detection angle $\phi_{HD}$, relative to the squeezing angle, is another free parameter. Building on the approach of Ref.~\cite{meyer2021variational}, we use the inverse of the classical Fisher information~\cite{meyer2021fisher} as a cost function $C(\phi_{\alpha}, \phi_{HD}) = {1}/{\mathcal{F}(\phi_{\alpha}, \phi_{HD})}$, where the inverse of the Fisher information represents the lowest achievable variance with many repetitions of the experiment, serving as a good proxy for metrological precision. The Supplementary Material (sections 2 and 3) details the computation of this metric in continuous variable systems using Gaussian parameter-shift rules. 
As outlined in the introduction, we combine this setup with two different optimizers: a gradient-based one for ab initio optimization and a Bayesian one for fine-tuning. 

\section{Results}
\subsection{Experimental principle}
The principle behind the experiment is illustrated in \cref{fig:principle}. We prepare a displaced squeezed state by pumping a hemilithic optical parametric oscillator (OPO) (previously described in \cite{Arnbak2019}) at \SI{775}{\nano\meter}. This process generates squeezed vacuum at \SI{1550}{\nano\meter}. The squeezed vacuum is then combined with a coherent state on a $99/1$ beamsplitter, resulting in the displacement of the squeezed light at the \SI{5}{\mega\hertz} sideband using a phase modulator in the coherent beam. After interacting with the lab environment, the squeezed light is measured by a homodyne detector. The relative phases between the squeezed light, the local oscillator and the displacement beam are locked using the coherent locking technique \cite{vahlbruch2016_coherent}. This technique, employing a \SI{40}{\mega\hertz} phase-locked sideband mode transmitted alongside the squeezed light, serves as a phase reference and allows full access to the phase space for both the squeezed light and displacement. 
 
A detailed description of the experimental setup can be found in the Supplementary Material fig. 1. For the measurements in this paper, the OPO's pump power is 2.7 mW resulting in the measurement of around 5 dB squeezing and 12 dB anti-squeezing. The displacement added to the squeezing is approximately $\alpha = 5.2$, a regime of interest since the contributions of the squeezed and the coherent photons to the classical Fisher Information are comparable.

The homodyne detector's output is sampled using a data-acquisition card after which it is downmixed to the \SI{5}{\mega\hertz} sideband and subsequently lowpass filtered with bandwidth \SI{1}{\mega\hertz}. The processed data are then used to characterize the measurement statistics.  
The cost function is estimated by varying the phases of the local oscillator according to the parameter-shift rules (see Supplementary Material section 2). Depending on the algorithm employed, the local oscillator and displacement phases are either further shifted during measurements to generate gradients for the gradient descent algorithm, or the cost function is directly fed to the Bayesian optimizer for the gradient-free algorithm. Based on either the gradient of the cost function or the decision function of the Bayesian optimizer, a new set of phase parameters is found. The process is repeated until optimization is achieved. 

\subsection{Gradient descent optimization}

\begin{figure*}
    \centering
    \includegraphics{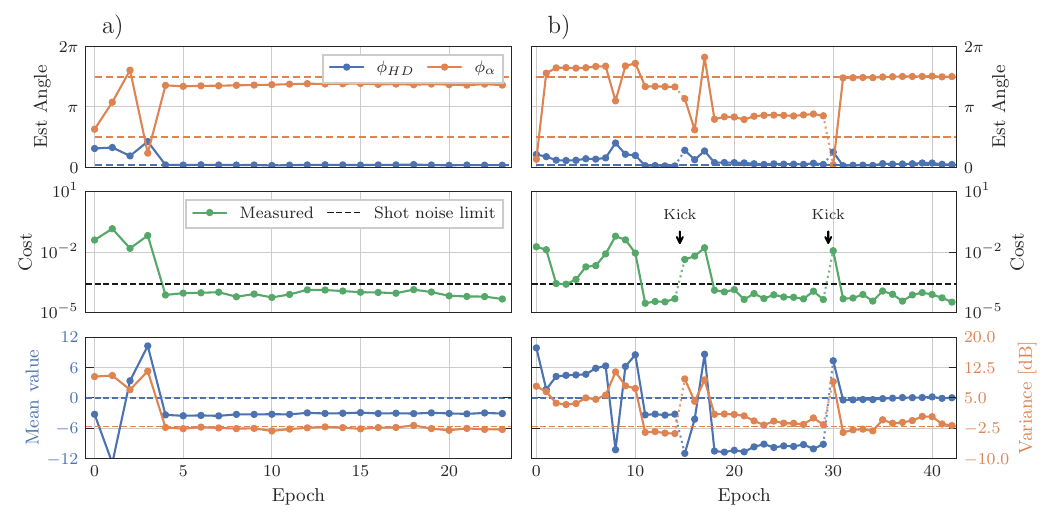}
    \caption{a) Demonstration of the gradient descent algorithm across 24 optimization epochs. (Top) The optical phase angles $\phi_{HD}$ and $\phi_\alpha$ estimated from the measurements. (Middle) The measured cost function $C = 1/\mathcal{F}$. The dotted line represents the shot noise limit accounting for the number of photons in the measurement and the number of samples used to estimate the cost function. (Bottom) The measured quadrature mean values and variances. The dotted lines in the top and bottom plots indicate the optimal values as predicted by theory (See Supplementary section 2). Note that in this particular measurement, the displacement appears to be slightly larger than $\alpha = 5.2$, likely due  to improved spatial overlap between the displacement beam and local oscillator, as the added modulation was consistent across all measurements. b) Kick-test of the gradient descent optimization over 45 epochs. The arrows labeled "kick" indicate the points immediately following the application of a kick.}
    \label{fig:Grad}
\end{figure*}

In \cref{fig:Grad} a), we present a run of the gradient descent-based optimization over 24 epochs. There is a clear trend of the cost function starting at a high value and then converging to a low one. The black dotted line of the middle plot of \cref{fig:Grad} a) represents the theoretical shot noise limit, which is based on the average number of photons in the probe state and the number of samples used to estimate each value of the cost function. Notably, the optimized cost function falls below this limit after convergence, indicating that the algorithm successfully identifies an optimum below the classical limit. Examining the resulting quadrature values, we observe the variance dropping below shot noise and approaching a value determined by the basis angle that maximizes the quadrature variance contribution to our analytical model of the classical Fisher Information. The theoretical model is elaborated upon in great detail in Supplementary Material sections 2 and 5. In general, it can be described as a squeezed, displaced probe state undergoing optical loss. This can be modelled analytically (see eq. 8 of the Supplementary Material). 

We observe unexpected local minima during the optimizer runs, evident from the mean value of the measurement quadrature. Ideally, we would expect this mean value to stabilize around 0, but it occasionally stabilizes around an intermediate value (in this case $\sim4$). This is not expected from the analytical model of our experiment (eq. 8 in the Supplementary Material).

We try to simulate the gradient descent experiment by numerically introducing Gaussian distributed phase noise to approximate the experimental conditions as closely as possible, and while we can observe a similar trend, the simulation does not find the exact same minimum, showing the difficulty in modelling experimental noise environments. The exact cause of this phenomenon is a subject for further investigation, but it does, however, showcase the advantage of variational approaches, as they can naturally uncover minima not predicted by either the analytical or the numerical model of our experiment.

In \cref{fig:Grad} b), we display the results of a different run of the experiment.
To test the algorithm's robustness against disturbances, we allowed the system to optimize for 14 rounds. Then, on the 15th round, we apply a kick to both control parameters, effectively displacing the system from its optimum. Subsequently, the system was permitted to optimize for another 14 epochs, before being subjected to another kick. 

Once again, we observe that the algorithm successfully optimizes the system after each kick, consistently reaching below the shot noise limit. While the quadrature variance converges towards the expected value, the mean value tends to stabilize around an intermediate point. Interestingly, in this particular round, during the final optimization step, the mean value does settle around the expected optimum. However, we also note that the difference in cost function between the various minima is very small and essentially indistinguishable in our measurements. This observation suggests that the optimization of the variance significantly impacts the cost function more than the mean value. This hypothesis is further supported by a simulation presented in fig. 9 of the Supplementary Material. This simulation, a 2D simulation of the cost-function landscape with the kick-measurements superimposed, reveals that the minima are quite broad and shallow as a function of the displacement angle, especially once the measurement angle has been optimized.    

\subsection{Post-hoc Bayesian optimization for fine-tuning}

In our search for an even better solution than those obtained from the gradient descent approach, we employ a post-hoc gradient-free optimization using data from low-cost areas. We use Bayesian optimization (BO), a probabilistic algorithm often used in machine learning for tuning model hyperparameters \cite{snoek2012practical,gortler2019visual,agnihotri2020exploring}. 
Bayesian optimization is particularly suited to our experiment as it utilizes a probabilistic approximation of the underlying model. This approach is beneficial for incorporating uncertainty related to our necessarily inexact modelling of the system and the to random processes occurring in the experiment.
Furthermore, the gradient-based algorithm requires five measurements to estimate the cost function and an additional eight measurements per control parameter to estimate the gradient, whereas the gradient-free algorithm only needs the five measurements for the cost function estimation. This results in a significant reduction in experimental resources per optimization step.

A Bayesian optimization routine comprises two parts. First, a surrogate function models the underlying cost function landscape $C(\phi_{HD},\phi_{\alpha})$ probabilistically, indicated by the predictive mean $\mu(C(\phi_{HD},\phi_{\alpha})) \approx C(\phi_{HD},\phi_{\alpha})$ and an uncertainty captured by the covariance matrix $\Sigma(C(\phi_{HD},\phi_{\alpha}))$. Second, an acquisition function determines which data points to query during iterative optimization -- essentially, it quantitatively identifies a ``good'' next point for the experiment. 

We selected a Gaussian Process (GP)\cite{rasmussen2003gaussian} for the surrogate function. For every point $(\phi_{HD},\phi_{\alpha})$, the predictive distribution 
of the surrogate model follows a normal distribution with mean $\mu(C(\phi_{HD},\phi_{\alpha}))$ and covariance $\Sigma(C(\phi_{HD},\phi_{\alpha}))$. The model's covariance is a key hyperparameter that balances the trade-off between exploration (favoring larger steps in parameter space) and exploitation (focusing on the vicinity of the current optimum).
For the acquisition function, we opted for Expected Improvement (EI), a commonly used technique in Bayesian optimization~\cite{frazier2018bayesian}. Intuitively, EI selects the next point that according to the current model is expected to achieve the lowest value of the cost function. 

\begin{figure*}
    \centering\includegraphics{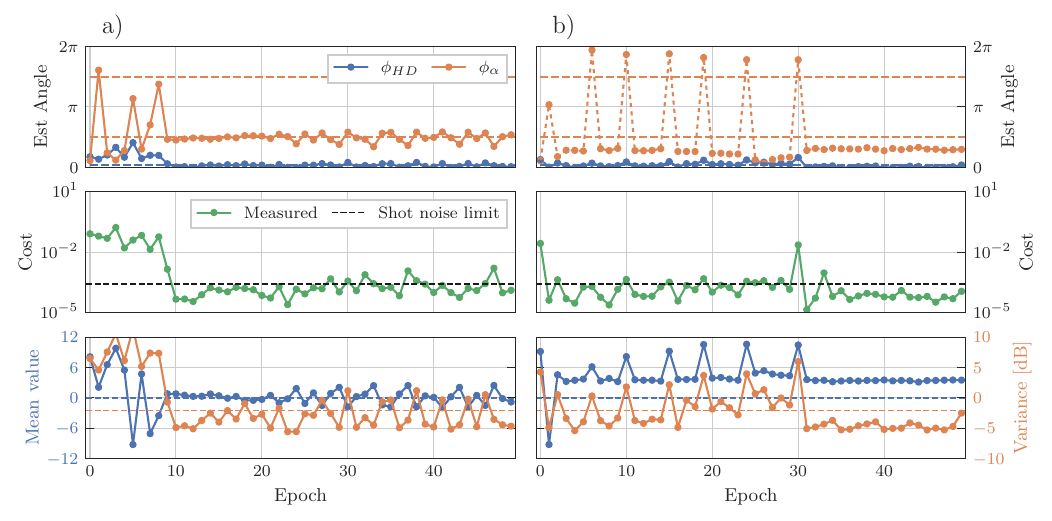}
    \caption{Demonstration of Bayesian optimization over 50 epochs, with a) loose and b) optimized hyper-parameters of the Gaussian Process (see Supplementary Material). (Top) The optical phase angles $\phi_{HD}$ and $\phi_\alpha$ estimated from the measurements. (Middle) The measured cost function $C = 1/\mathcal{F}$. The dotted line represents the shot noise limit, accounting for the number of photons in the measurement and the number of samples used to estimate the cost function. (Bottom) The measured quadrature mean values and variances. The dotted lines in the top and bottom plots indicate the optimal values as predicted by theory (Supplementary section 2).}
    \label{fig:NoGrad_long}
\end{figure*}

Initially, the Gaussian Process was fitted with a data set of 136 data points obtained from the gradient descent experiment along with randomly sampled points in our search space. 
This approach is hence referred to as \textit{post-hoc} Bayesian optimization for fine-tuning. The next parameter set for the experiment is determined by maximizing the acquisition function, which relies solely on the surrogate model of the Bayesian Optimizer. 

After selecting this new set of parameters, the physical experiment is conducted, and the actual cost value is obtained. This new data point is then incorporated into the dataset, and the updated dataset is used to fit a new Gaussian Process for the subsequent step of the algorithm. 
This process is repeated iteratively for 50 epochs, and report the lowest obtained cost function value. 
See the Supplementary Material for technical details and Sec. 2.6--2.8 in~\cite{foldager2023quantum} for a full derivation.

From \cref{fig:NoGrad_long}, we observe once again that the algorithm reaches an optimum of the system close to the expected theoretical minimum. In part a) of \cref{fig:NoGrad_long}, we set the hyperparameters of the Gaussian Process more loosely, resulting in the algorithm exploring the parameter landscape around the minimum. This exploration comes at the cost of optimization process's stability, causing the cost function to fluctuate to relatively high values. Optimizing the balance between exploration and exploitation is achievable by tuning the hyper-parameters of the model. The results of this tuning are evident in part b) of \cref{fig:NoGrad_long}, where the model initially focuses on exploring new areas of cost function landscape $C(\phi_{HD},\phi_{\alpha})$ for the first 30 epochs. It then shifts to a more stable exploitation of the achieved minimum in the subsequent 20 epochs. Although not depicted in this trace, the gradient-free algorithm was also exhibited susceptibility to the same local minima as those identified using the gradient descent-based algorithm. This adds further evidence that our theoretical model does not fully capture the entire noise landscape of the experiment.

The gradient-free Bayesian optimization method employed here offers two advantages over the gradient descent optimization. First, it reduces  the risk of getting trapped in a local minimum, a common problem with gradient descent. Second, as it does not require the evaluation of the gradient, it uses less experimental resources.  
However, employing Bayesian optimization is not without its drawbacks. As discussed in previous sections, the gradient-based method will be able to follow slow drifts of the system as these drifts alter the gradients. In contrast, the Bayesian Optimizer is likely to be slower in optimizing a non-stationary system due to the need for an exploration phase whenever the experimental setup undergoes changes. Additionally, the performance of the gradient-free algorithm heavily depends on the careful optimization of its hyperparameters.

\section{Discussion}

In summary, to our knowledge, this is the first instance of testing hybrid quantum-classical optimization algorithms for optical phase estimation with squeezed coherent light and homodyne detection. Specifically, we have investigated the performance of both a gradient-based optimization and post-hoc gradient-free Bayesian optimization. 

Our results confirm that both algorithms can successfully adjust the control parameters of the system to achieve optimal estimation performance. This includes preparing of the optimal probe state and setting the measurement parameters. Notably, the algorithms achieve this without prior knowledge about the noise processes in the hardware, as evidenced by the discovery of optima that were not anticipated by our theoretical model. Additionally, we have shown how the gradient-based algorithm can automatically adjust the system to the optimal setting when the phase to be estimated changes. 

These findings underscore the potential of variational quantum algorithm-based quantum metrology, particularly for optical phase estimation with squeezed coherent light and homodyne detection. This motivates further investigation of such techniques in more complex quantum sensing systems with additional control parameters, such as the degree of squeezing and the level of coherent excitation. In addition, the application of these variational algorithms in multi-parameter sensing systems~\cite{Guo2020}, which are relevant for quantum imaging~\cite{Kolobov2007}, entangled sensor networks~\cite{Brady2022}, and networked atomic clocks~\cite{Komar2014}, is also a promising avenue.  

Our study also reveals that the choice of classical optimizers may vary depending on the specific estimation task. We observe that the gradient-based optimizer is effective for ab initio phase estimation but incurs a higher measurement overhead compared to the gradient-free Bayesian optimization. Conversely, our implementation of the Bayesian optimization was adept at fine-tuning control parameters, though it required a training set from the gradient descent optimization for optimal performance. We also noted that careful optimization of the hyper-parameters of the algorithms such as the learning rate and ratio of exploration/exploitation is crucial for good performance. Further investigations in this area, including the potential benefit of switching between different classical optimizers for practical phase estimation tasks, will be vital to further validate the practical applicability of variational techniques in quantum-enhanced metrology.
\section{Acknowledgements}
J.N., M.K., T.A., J.N.-N., T.G. and U.A. acknowledge support from the Danish
National Research Foundation, Center for Macroscopic
Quantum States (bigQ, DNRF142).
K.V. acknowledges the European Union's Horizon 2020 research and innovation programme under the Marie Sklodowska-Curie Grant Agreement No. 956071 (AppQInfo). J.B. acknowledges funding and support from the NWO Gravitation Program Quantum Software Consortium (Project QSC No. 024.003.037) and The AWS Quantum Discovery Fund at the Harvard Quantum Initiative.

\bibliographystyle{ieeetr}
\bibliography{Ref_PhaseEstimation}

\end{document}


\title{Variational quantum algorithm for enhanced continuous variable optical phase sensing: Supplementary Material}
\maketitle
\section{Experimental System}
\begin{figure*}[htp]
    \centering
    \includegraphics[width = \linewidth]{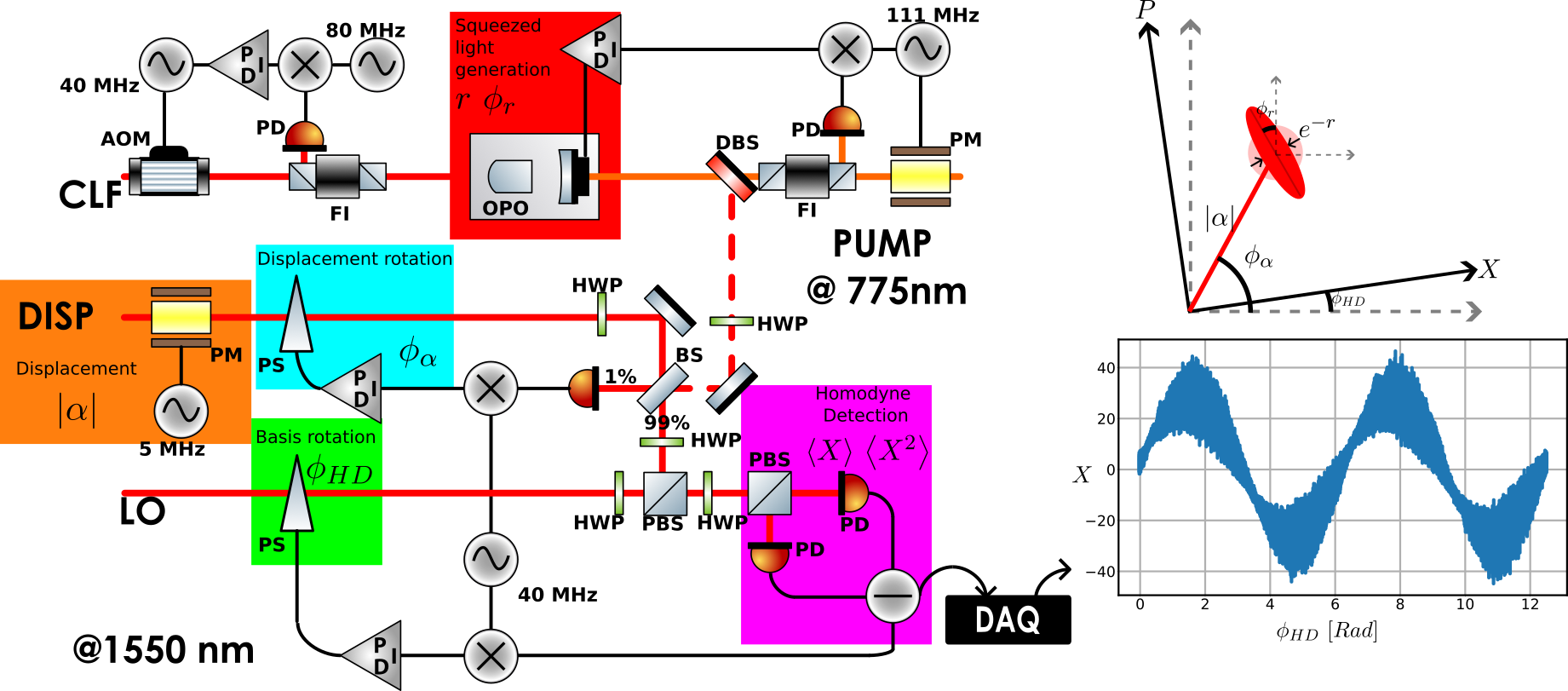}
    \caption{Sketch of the experimental setup. Abbreviations: CLF: Coherent-locking-field, DISP: Displacement field, LO: Local oscillator, DAQ: Data acquisition. OPO: Optical-parametric-oscillator. PD: Photo-detector. FI: Faraday Isolator. BS: Beam-splitter. AOM: Acousto-optic oscillator. PM: Phase modulator. PS: Phase shifter. PBS: Polarizing beam-splitter. DBS: Dichroic beam-splitter. HWP: Half-wave plate.}
    \label{fig: setup}
\end{figure*}

The experimental system is shown in figure \cref{fig: setup}. A hemilithic, double resonant, optical-parametrtic-oscillator (OPO) is pumped with a light field at \SI{775}{\nano\meter}, and the squeezed light is generated at \SI{1550}{\nano\meter}. The squeezed light source, which is a modification of the source presented in \cite{Nielsen2023}, has a FWHM bandwidth of \SI{66}{\mega\hertz} and a threshold power of around \SI{6}{\milli\watt}. The source is pumped with \SI{2.7}{\milli\watt} and the system has around \SI{72}{\percent} efficiency and \SI{30}{\milli\radian} phase noise RMS (between squeezed light and local oscillator), measuring around 5 dB of squeezed light and 11.8 dB of anti-squeezed light at a sideband frequency of \SI{5}{\mega\hertz}. From these a squeezing strength $r \sim 1.52$ can be estimated. 

The OPO is stabilized via the Pound-Drever-Hall (PDH) technique \cite{Drever1983}, and a \SI{40}{\mega\hertz} frequency-shifted beam is injected into the OPO to act as a phase reference in a coherent locking scheme \cite{vahlbruch2016_coherent}. This \SI{40}{\mega\hertz} reference tone (CLF) is used to stabilize the phase between the squeezed light, the displacement beam (DISP) and the local oscillator (LO). Tuning the electrical down-mixing phase of the two phase locks allows full control of the displacement angle $\phi_{\alpha}$ and the homodyne basis angle $\phi_{HD}$ relative to the squeezed quadrature angle $\phi_r$, which is arbitrarily set by the phase of the pump light. 

\section{Classical Fisher Information of a Gaussian state and the cost function.}
\label{sec:CFisher}

 The cost function used in the experiment is the inverse of the classical Fisher Information
 \begin{align}
 C = \frac{1}{\mathcal{F}}
 \end{align}
 We assume our system to be a CW Gaussian state described by quadrature operators $\vu{X} \equiv (\vu{a}^\dagger + \vu{a})$ and $\vu{P} \equiv i(\vu{a}^\dagger - \vu{a})$ and fully characterized by the first two statistical moments $\mu_x = \expval{\vu{X}}$ and $V_x = \expval{\vu{X}^2}-\expval{\vu{X}}^2$ (with similar moments for $\vu{P}$) and with a  photon number operator given by
 \begin{align}        
\vu{n} = \frac{1}{4}\qty(\vu{X}^2+\vu{P}^2-2\mathbb{I}),
 \end{align}
  using the commutator $\comm{\vu{X}}{\vu{P}} = 2i$.
  
  The classical Fisher Information for the estimation of phase shifts $\phi$, with a measurement probability distribution of $\vu{X}$ quadrature values $P(x)$ parameterized by parameters $\qty{\Theta}$ is given by
 \begin{align}
 \mathcal{F} = \bigintss \dd x P\qty(x|\phi,\qty{\Theta})\qty(\pdv{\log(P\qty(x|\phi, \qty{\Theta}))}{\phi})^2. \label{eq: genF}
 \end{align}    
\Cref{eq: genF} can be evaluated, since we have a Gaussian state, as
\begin{align}
\mathcal{F} = \frac{1}{V_x}\qty(\pdv{\mu_x}{\phi})^2 + \frac{1}{2 V_x^2}\qty(\pdv{V_x}{\phi})^2,
\label{eq: gaussF}
\end{align}
where $\mu_x$ and $V_x$ will be functions of the parameters $\qty{\Theta}$.

The gradient of the cost function with respect to the parameters $\Theta$ is given by
\begin{align}
\grad_\Theta{C} = \frac{-\grad_\Theta{\mathcal{F}}}{\mathcal{F}^2} .
\end{align}
The gradient of the classical Fisher Information is finally given by
\begin{align}
\grad_\Theta{\mathcal{F}} &= \frac{1}{V^3_x}\qty(2V_x^2\pdv{\mu_x}{\phi}\grad_\Theta{\pdv{\mu_x}{\phi}}-V_x\qty(\pdv{\mu_x}{\phi})^2\grad_\Theta{V_x})
+\frac{1}{V^3_x}\qty(V_x\pdv{V_x}{\phi}\grad_\Theta{\pdv{V_x}{\phi}}-\qty(\pdv{V_x}{\phi})^2\grad_\Theta{V_x}).
\end{align}

The derivative of the variance is in general given by
\begin{align}
    \pdv{V_x}{\phi}=\pdv{}{\phi}\expval{\vu{X}^2}-2\expval{\vu{X}}\pdv{}{\phi}\expval{\vu{X}} 
\end{align}.
The two main contributions to the loss of information are the loss of squeezed photons and phase noise between the three interacting fields (displacement- and squeezing fields and local oscillator). The loss of photons comes from the limited escape efficiency of the squeezer $\eta_{esc}$, the limited efficiency of optical components between the squeezer and the detector $\eta_{opt}$, the imperfect visibility between signal and local oscillator $\mathcal{V}^2$ and finally the imperfect quantum efficiency of the photodiodes $\eta_{QE}$. 
Phase noise comes mainly from the inability of the the phase-stabilization loops to remove all classical phase fluctuations due to having limited bandwidths and shot-noise of the light fields.
In the case of pure loss, ignoring phase noise, the Classical Fisher Information can be evaluated analytically by starting from \cref{eq: gaussF}
\begin{align}
    \mathcal{F}(\phi) &= \frac{1}{V_x}\qty(\pdv{\mu_x}{\phi})^2 + \frac{1}{2 V_x^2}\qty(\pdv{V_x}{\phi})^2\nonumber\\
    &= \frac{1}{\eta\qty(e^{-2r}\cos[2](\phi)+e^{2r}\sin[2](\phi))+1-\eta}\qty(\pdv{}{\phi}2\sqrt{\eta}\abs{\alpha}\cos(\phi_\alpha-\phi))^2\nonumber\\
    &+\frac{1}{\qty(\eta\qty(e^{-2r}\cos[2](\phi)+e^{2r}\sin[2](\phi))+1-\eta)^2} \qty(\pdv{}{\phi}\qty[\eta\qty(e^{-2r}\cos[2](\phi)+e^{2r}\sin[2](\phi))+1-\eta])^2 \nonumber\\
    &=\frac{4\eta\abs{\alpha}^2\sin[2](\phi_\alpha-\phi)}{\eta\qty(e^{-2r}\cos[2](\phi)+e^{2r}\sin[2](\phi))+1-\eta}
    +\frac{2\eta^2\sinh[2](2r)\sin[2](2\phi)}{\qty(\eta\qty(e^{-2r}\cos[2](\phi)+e^{2r}\sin[2](\phi))+1-\eta)^2}. \label{eq: FullFisher}
\end{align}
In the zero loss scenario, the variance term will always scale faster with increasing photon numbers, meaning that concentrating the photons in the squeezed state is the more effective strategy. In this case the optimal measurement phase is given by $\phi_{opt} = \arccos(\tanh(2r))/2$.\\
In the case of relatively low loss, there exists an optimal ratio between squeezed and coherent photons with an optimal relative angle being $\phi_\alpha-\phi = \pi/2$ and an optimal measurement angle given by
\begin{align}
\phi_{opt}^\mathcal{L} = \frac{1}{2}\arccos(\frac{\frac{\eta e^{2r}+(1-\eta)}{\eta e^{-2r}+(1-\eta)}-1}{\frac{\eta e^{2r}+(1-\eta)}{\eta e^{-2r}+(1-\eta)}+1}).
\end{align}
In the high loss scenario, the mean value term will dominate meaning most photons should be put into the coherent state, with the optimal relative angle being $\phi_\alpha-\phi = \pi/2$ and the optimal measurement angle being $\phi = 0$.

\section{Parameter-shift rules for quadratic operators.}

In general, an arbitrary operator can be expressed as 
\begin{align}
\vu{A} = \mqty(a, b, c, d, e \hdots)\cdot\mqty(\mathbb{I} \\ \vu{X} \\ \vu{P} \\ \vu{X}^2 \\ \vu{P}^2\\ \vdots)
\end{align}
When a gate acts upon this arbitrary operator $\vu{A}$, we in principle need to know the infinite-dimensional gate matrix describing the transformation of all entries of the vector $\vu{A}_G = G[\vu{A}] = M_G^T\vu{A}$, where $M_G$ is a matrix describing the action of the gate upon the operator. Please note the transposition of the matrix - this is a very subtle and important detail. The derivatives of operators linear in quadratures (e.g. $\grad_\Theta{\expval{\vu{X}}}$) can easily be evaluated using the parameter shift rules from \cite{Schuld2019}.
to find the higher-order entries of the gate matrix or truncate the vector space to only include the linear operators. It is also possible to derive parameter shift rules for operators quadratic in the quadratures, by either finding and differentiating the higher order entries of the gate matrices using
\begin{align}
G[\vu{A}\vu{B}] = G[\vu{A}]G[\vu{B}] \label{eq: higherorder},
\end{align}
or 
\begin{align}
\pdv{}{\Theta_i}G(\Theta_i)\qty[\vu{A}^2] &=
\pdv{}{\Theta_i}G(\Theta_i)\qty[\vu{A}]G(\Theta_i)\qty[\vu{A}]+ G(\Theta_i)\qty[\vu{A}]\pdv{}{\Theta_i}G(\Theta_i)\qty[\vu{A}], \label{eq: highquad} 
\end{align}
where $G[\Theta_i]$ is a gate parameterized by the parameter $\Theta_i$, and where \cref{eq: highquad} has been truncated to only include linear quadrature operators $\vu{A} \subseteq [\mathbb{I},\vu{X},\vu{P}]$.

In the following, we will reproduce the results of \cite{Schuld2019}, finding the linear parameter shift rules for the relevant Gaussian gates. We will then extend this to also include quadratic operators. Throughout this we will adopt the definition of the Gaussian gates used in \cite{Schuld2019}. 
\subsection{Squeezing gate for linear operators}
The squeezing gate is parameterized by $r$ the squeezing strength and the gate matrix for linear operators is given by
\begin{align}
M_S(r) &= \mqty(\dmat[0]{1,e^{-r},e^r}),\\
\pdv{}{r}M_S(r) & = \mqty(\dmat[0]{0,-e^{-r},e^r}).
\end{align}
We now want to express the derivate of the matrix as a linear superposition of the gate matrix itself
\begin{align}
\pdv{}{r}M_S(r) & = \mqty(\dmat[0]{0,-e^{-r}\frac{e^s-e^{-s}}{2\sinh(s)},e^r\frac{e^s-e^{-s}}{2\sinh(s)}})\nonumber \\
 & =\frac{1}{2\sinh(s)} \mqty(\dmat[0]{0,e^{-(r+s)}-e^{-(r-s)},e^{(r+s)}-e^{(r-s)}})\nonumber\\
&= \frac{1}{2\sinh(s)}\qty(M_S(r+s)-M_S(r-s)),
\end{align}
which is the parameter shift rule for the squeezing gate, with s being an arbitrary shift in the squeezing strength.
\subsection{Displacement gate for linear operators}
The displacement gate is parameterized by $\alpha$ the displacement amplitude and $\phi_\alpha$ the displacement angle and given by the matrix
\begin{align}
M_D(\alpha, \phi_\alpha) &= \mqty(1 & 0 & 0 \\ 2\alpha \cos(\phi_\alpha) & 1 & 0 \\ 2\alpha \sin(\phi_\alpha) & 0 & 1),\\
\pdv{}{\alpha}M_D(\alpha, \phi_\alpha) & = \mqty(0 & 0 & 0 \\ 2 \cos(\phi_\alpha) & 0 & 0 \\ 2 \sin(\phi_\alpha) & 0 & 0),\\
\pdv{}{\phi_\alpha}M_D(\alpha, \phi_\alpha) & = \mqty(0 & 0 & 0 \\ -2\alpha \sin(\phi_\alpha) & 0 & 0 \\ 2\alpha\cos(\phi_\alpha) & 0 & 0).
\end{align}
We can once again easily calculate the parameter shift rule for the linear operators
\begin{align}
\pdv{}{\alpha}M_D(\alpha, \phi_\alpha)  &= \mqty(0 & 0 & 0 \\ 2 \cos(\phi_\alpha)\frac{\alpha+s-\qty(\alpha-s)}{2s} & 0 & 0 \\ 2 \sin(\phi_\alpha)\frac{\alpha+s-\qty(\alpha-s)}{2s} & 0 & 0)= \frac{1}{2s}\qty(M_D(\alpha+s, \phi_\alpha)-M_D(\alpha-s, \phi_\alpha)),\\
\pdv{}{\phi_\alpha}M_D(\alpha, \phi_\alpha) &= 
\mqty(0 & 0 & 0 \\ 2\alpha \frac{1}{2}\qty(\cos(\phi_\alpha+\pi/2)-\cos(\phi_\alpha-\pi/2)) & 0 & 0 \\ 2\alpha\frac{1}{2}\qty(\sin(\phi_\alpha+\pi/2)-\sin(\phi_\alpha-\pi/2)) & 0 & 0)= \frac{1}{2}\qty(M_D(\alpha, \phi_\alpha+\pi/2)-M_D(\alpha, \phi_\alpha-\pi/2)).
\end{align} 
\subsection{Rotation gate for linear operators}
The final gate of this analysis is the rotation gate $R(\phi)$ given by the matrix 
\begin{align}
M_R(\phi) &= \mqty(1 & 0 & 0 \\ 0 & \cos(\phi) & -\sin(\phi) \\0 & \sin(\phi) & \cos(\phi)),\\
\pdv{}{\phi}M_R(\phi) &= \mqty(0 & 0 & 0 \\ 0 & -\sin(\phi) & -\cos(\phi) \\0 & \cos(\phi) & -\sin(\phi)),\\
\end{align}

From the derivative matrix we can (similar to the displacement gate) find the basic parameter shift rules by recognizing the linear difference of cosines shifted by $\pi/2$ gives sine and vice versa, and we arrive at the same parameter shift rules as with the displacement angle

\begin{align}
\pdv{}{\phi}M_R(\phi) &= \frac{1}{2}\qty(M_R(\phi+\pi/2)-M_R(\phi-\pi/2)).
\end{align}
\subsection{Squeezing gate for quadratic operators}
We begin by applying \cref{eq: higherorder} to the squeezing gate to find the higher-order entries of the gate matrix
\begin{align}
S[\vu{X}\vu{X}] &= S[\vu{X}]S[\vu{X}] = e^{-2r} \vu{X}^2,\\
S[\vu{P}\vu{P}] &= S[\vu{P}]S[\vu{P}] = e^{2r} \vu{P}^2,\\
S[\vu{X}\vu{P}] &= S[\vu{X}]S[\vu{P}] = \vu{X}\vu{P},\\
S[\vu{P}\vu{X}] &= S[\vu{P}]S[\vu{X}] = \vu{P}\vu{X}.
\end{align}
The  gate matrix is then expanded to include
\begin{align}
M_S(r) = \mqty(\dmat[0]{1,e^{-r},e^r,e^{-2r},e^{2r},1,1}).
\end{align}
If we truncate the matrix to only look at the quadratic terms, then we can derive parameters shift rules that apply to the quadratic operators
\begin{align}
\pdv{}{r}M_{S,quad}(r) &= \mqty(\dmat[0]{-2re^{-2r},2re^{2r},0,0})= \frac{1}{\sinh(2s)}\qty(M_{S,quad}(r+s)-M_{S,quad}(r-s))
\end{align}
\subsection{Displacement gate for quadratic operators}
Once again we can repeat the calculation from before, finding the entries of the gate matrix for the quadratic operators
\begin{align}
D(\alpha, \phi_\alpha)[\vu{X}\vu{X}] &= (2\alpha\cos(\phi_\alpha)+\vu{X})(2\alpha\cos(\phi_\alpha)+\vu{X})  = 4\alpha^2\cos[2](\phi_\alpha)+\vu{X}^2+4\alpha\cos(\phi_\alpha)\vu{X},\\
D(\alpha, \phi_\alpha)[\vu{P}\vu{P}] &= (2\alpha\sin(\phi_\alpha)+\vu{P})(2\alpha\sin(\phi_\alpha)+\vu{P}) = 4\alpha^2\sin[2](\phi_\alpha)+\vu{P}^2+4\alpha\sin(\phi_\alpha)\vu{P},\\
D(\alpha, \phi_\alpha)[\vu{X}\vu{P}] &= (2\alpha\cos(\phi_\alpha)+\vu{X})(2\alpha\sin(\phi_\alpha)+\vu{P}) = 4\alpha^2\sin(2\phi_\alpha)+\vu{X}\vu{P}+2\alpha\qty(\cos(\phi_\alpha)\vu{P}+\sin(\phi_\alpha)\vu{X}),\\
D(\alpha, \phi_\alpha)[\vu{P}\vu{X}] &= (2\alpha\sin(\phi_\alpha)+\vu{P})(2\alpha\cos(\phi_\alpha)+\vu{X}) = 4\alpha^2\sin(2\phi_\alpha)+\vu{P}\vu{X}+2\alpha\qty(\cos(\phi_\alpha)\vu{P}+\sin(\phi_\alpha)\vu{X}).
\end{align}
The resulting gate matrix including quadratic operators is then
\begin{align}
M_D(\alpha, \phi_\alpha) = \mqty(1 & 0 & 0 & 0 & 0 & 0 & 0 \\ 2\alpha \cos(\phi_\alpha) & 1 & 0 & 0 & 0 & 0 & 0  \\ 2\alpha \sin(\phi_\alpha) & 0 & 1 & 0 & 0 & 0 & 0 \\
4\alpha^2 \cos[2](\phi_\alpha) & 4\alpha\cos(\phi_\alpha) & 0 & 1 & 0 & 0 & 0 \\ 4\alpha^2 \sin[2](\phi_\alpha) & 0 & 4\alpha\sin(\phi_\alpha) & 0 & 1 & 0 & 0 \\
4\alpha^2 \sin(2\phi_\alpha) & 2\alpha\sin(\phi_\alpha) & 2\alpha\cos(\phi_\alpha) & 0 & 0 & 1 & 0  \\ 4\alpha^2 \sin(2\phi_\alpha) & 2\alpha\sin(\phi_\alpha) & 2\alpha\cos(\phi_\alpha) & 0 & 0 & 0 & 1 ) .
\end{align}
Once again, we limit ourselves to only quadratic operators and find the corresponding parameter shift rules. This is a bit more involved than with the squeezing gate, but if we consider the first column with terms proportional to $\alpha^2$. For the displacement amplitude $\alpha$, if we assume naively that the form of the parameter shift is the same as the linear one but with a different normalization, then differentiating gives us the following equation
\begin{align}
\pdv{4\alpha^2\cos[2](\phi_\alpha)}{\alpha} &= 8\alpha\cos[2](\phi_\alpha)= \frac{4}{k}\qty(\qty(\alpha+s)^2-\qty(\alpha-s)^2)\cos[2](\phi_\alpha)\Rightarrow\nonumber\\
2\alpha &= \frac{1}{k}\qty(\qty(\alpha+s)^2-\qty(\alpha-s)^2) \Rightarrow \nonumber\\
k &= 2s,
\end{align}
which leads to the same parameter shift rule as with the linear operators
\begin{align}
&\pdv{}{\alpha}M_{D,quad}(\alpha, \phi_\alpha) = \frac{1}{2s}\qty(M_{D,quad}(\alpha+s,\phi_\alpha)-M_{D,quad}(\alpha-s,\phi_\alpha)).
\end{align}
The calculation for the displacement angle $\phi_\alpha$ can be calculated by considering transformations of $\cos[2](\phi) \rightarrow -\sin(2\phi)$, $\sin[2](\phi) \rightarrow \sin(2\phi)$ and $\cos(\phi)\sin(\phi) \rightarrow \cos(2\phi)$. These results can be expressed by linear differences of the original functions shifted up and down by $\pi/4$ similar to the basic rotation gate parameter shift. The resulting parameter shift is then
\begin{align}
&\pdv{}{\phi_\alpha}M_{D,quad}(\alpha, \phi_\alpha) = \nonumber\\ &\qty(M_{D,quad}(\alpha,\phi_\alpha+\pi/4)-M_{D,quad}(\alpha,\phi_\alpha-\pi/4))+\qty(1-\frac{\sqrt{2}}{2})\qty(M_{D,quad}(\alpha,\phi_\alpha-\pi/2)-M_{D,quad}(\alpha,\phi_\alpha-\pi/2)).
\end{align}

The above expressions can be verified by looking at the derivatives of the number operator expectation value $\partial_\alpha \expval{\vu{n}} = 2\alpha$ and $\partial_{\phi\alpha} \expval{\vu{n}} = 0$, as we would expect.
\newline
\subsection{Rotation Gate for quadratic operators}
We begin again by finding the quadratic entries of the rotation matrix
\begin{align}
R(\phi)[\vu{X}\vu{X}] & =\cos[2](\phi)\vu{X}^2+\sin[2](\phi)\vu{P}^2-\cos(\phi)\sin(\phi)\qty(\vu{X}\vu{P}+\vu{P}\vu{X}),\\
R(\phi)[\vu{P}\vu{P}] & =\cos[2](\phi)\vu{P}^2+\sin[2](\phi)\vu{X}^2+\cos(\phi)\sin(\phi)\qty(\vu{X}\vu{P}+\vu{P}\vu{X}),\\
R(\phi)[\vu{X}\vu{P}] & =\cos[2](\phi)\vu{X}\vu{P}-\sin[2](\phi)\vu{P}\vu{X}+\cos(\phi)\sin(\phi)\qty(\vu{X}^2-\vu{P}^2),\\
R(\phi)[\vu{P}\vu{X}] & = \cos[2](\phi)\vu{P}\vu{X}-\sin[2](\phi)\vu{P}\vu{X}+\cos(\phi)\sin(\phi)\qty(\vu{X}^2-\vu{P}^2).
    \end{align}
    
    The gate matrix for rotations including quadratic operators is then given by
 \begin{align}
M_{R,quad}(\phi) =\mqty(1 & 0 & 0 & 0 & 0 & 0 & 0\\0 & \cos(\phi) & -\sin(\phi) & 0 & 0 & 0 & 0\\ 0 & -\sin(\phi) & \cos(\phi) & 0 & 0 & 0 & 0\\ 0 & 0 & 0 & \cos[2](\phi) & \sin[2](\phi) & -\cos(\phi)\sin(\phi) &-\cos(\phi)\sin(\phi)\\ 0 & 0 & 0 & \sin[2](\phi) & \cos[2](\phi) & \cos(\phi)\sin(\phi) & \cos(\phi)\sin(\phi)\\0 & 0 & 0 & \cos(\phi)\sin(\phi) & -\cos(\phi)\sin(\phi)  &\cos[2](\phi) & \sin[2](\phi)\\ 0 & 0 & 0 & \cos(\phi)\sin(\phi) & -\cos(\phi)\sin(\phi)  &\sin[2](\phi) & \cos[2](\phi)).
\end{align}

\begin{figure}
    \centering
    \includegraphics[width = 0.6\linewidth]{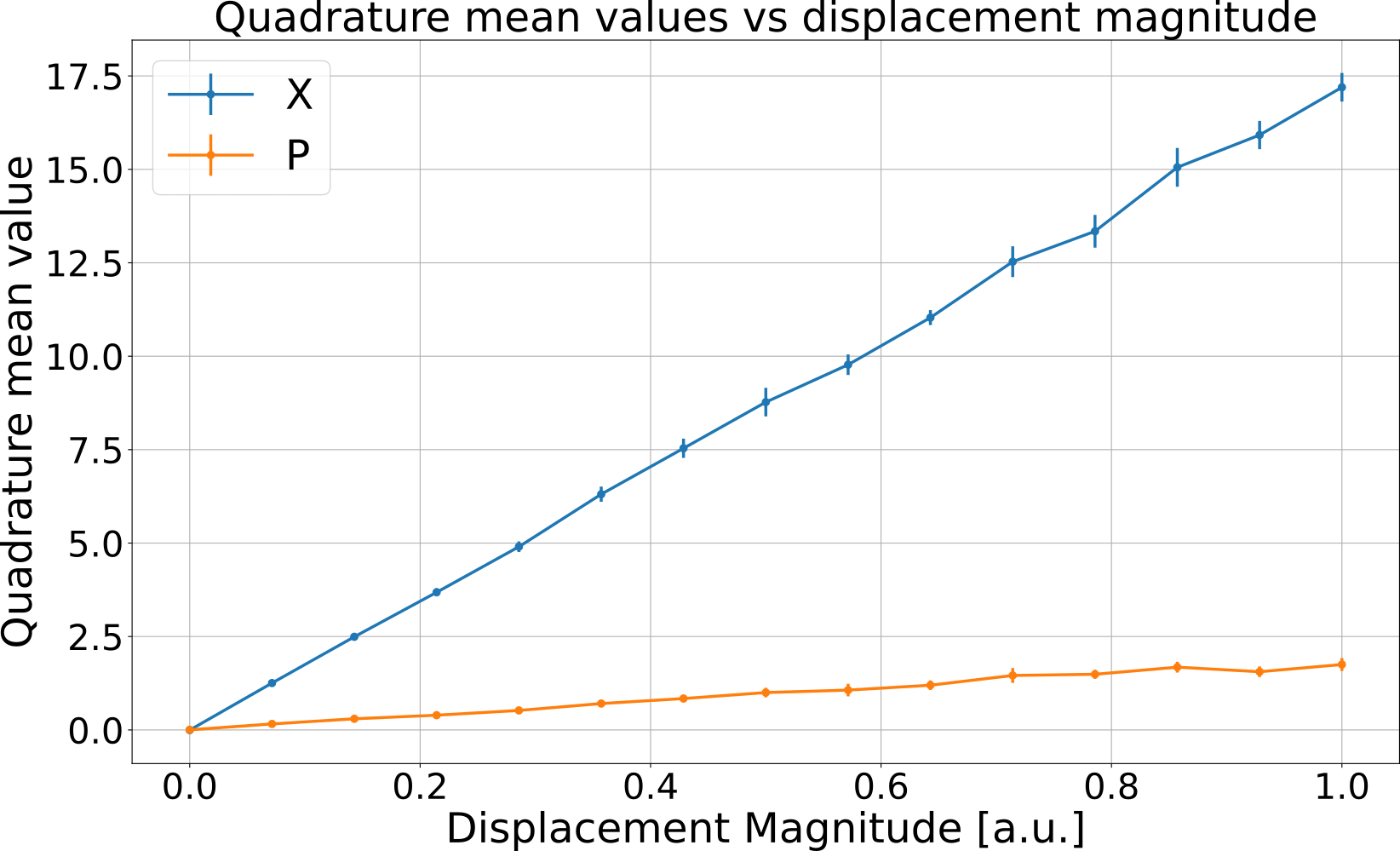}
    \caption{Plot of the X-and P quadrature mean values as a function of the function generator drive-magnitude at an arbitrary displacement phase $\phi_\alpha$.}
    \label{fig:alphacali}
\end{figure}

Once again differentiating $M_{R,quad}(\phi)$ is calculated using the same considerations as with $\phi_\alpha$, and we find it to be
\begin{align}
\pdv{}{\phi}M_{R,quad}(\phi) = \qty(M_{R,quad}(\phi+\pi/4)-M_{R,quad}(\phi-\pi/4)).
\end{align}

\begin{figure}
    \centering
    \includegraphics[width = 0.6\linewidth]{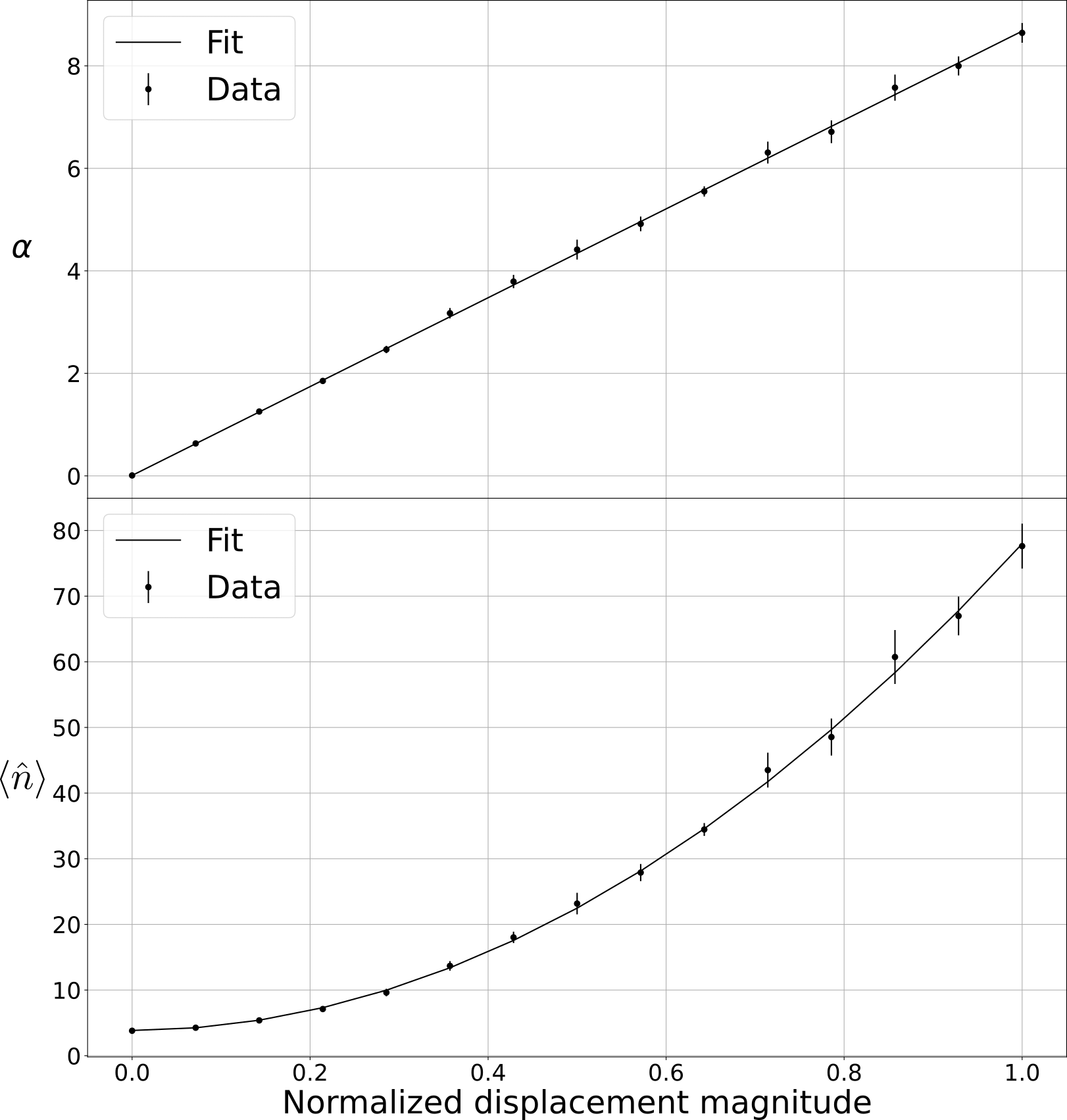}
    \caption{Plot of the estimated $\alpha$ and $\expval{n}$ as a function function generator drive magnitude. The zero-magnitude photon contribution comes from the squeezed state. The solid lines are linear and quadratic fits respectively.}
    \label{fig:alpha_n}
\end{figure}

\section{Calibrations}
Since the parameter shift rules assume a certain shift of experimental parameters, it is necessary to calibrate the experimental apparatus to make sure the correct operations are implemented. This section will deal with the calibrations of the displacement angle and magnitude and the homodyne angle.
\subsection{Displacement magnitude}

In the system, the displacement magnitude is set by the magnitude of the electronic oscillator driving the phase modulator that creates displaced photons at a sideband frequency of \SI{5}{\mega\hertz}. This is driven by a home-built AD9959 function generator with a Qudi\cite{BINDER2017} interface that sets the magnitude on an arbitrary scale between 0 and 1. To calibrate the displacement magnitude the function generator magnitude is swept and for each value the X-and P quadratures are sampled for \SI{30}{\milli\second} at a sampling rate of \SI{50}{\mega\hertz} and downmixed to the \SI{5}{\mega\hertz} sideband using a \SI{1}{\mega\hertz} lowpass filter. The data is normalized to the shotnoise standard deviation, and the mean values and squared mean values are calculated, and the measurement is repeated 10 times. $\phi_\alpha$ and $\phi_{HD}$ are arbitrary in this measurement The result of this measurement is shown in \cref{fig:alphacali}.

From this measurement, the displacement magnitude and photon number expectation value can be estimated as $\alpha = 1/2\sqrt{\expval{X}^2+\expval{P}^2}$ and $\expval{n} = 1/4\qty(\expval{X^2}+\expval{P^2}-2)$. These estimates are shown in \cref{fig:alpha_n}.

The plots also contain a linear fit of the displacement magnitude and a quadratic fit of the photon number. The non-zero photon expectation value at zero displacement comes from the squeezed state.

\begin{figure}
    \centering
    \includegraphics[width = 0.6\linewidth]{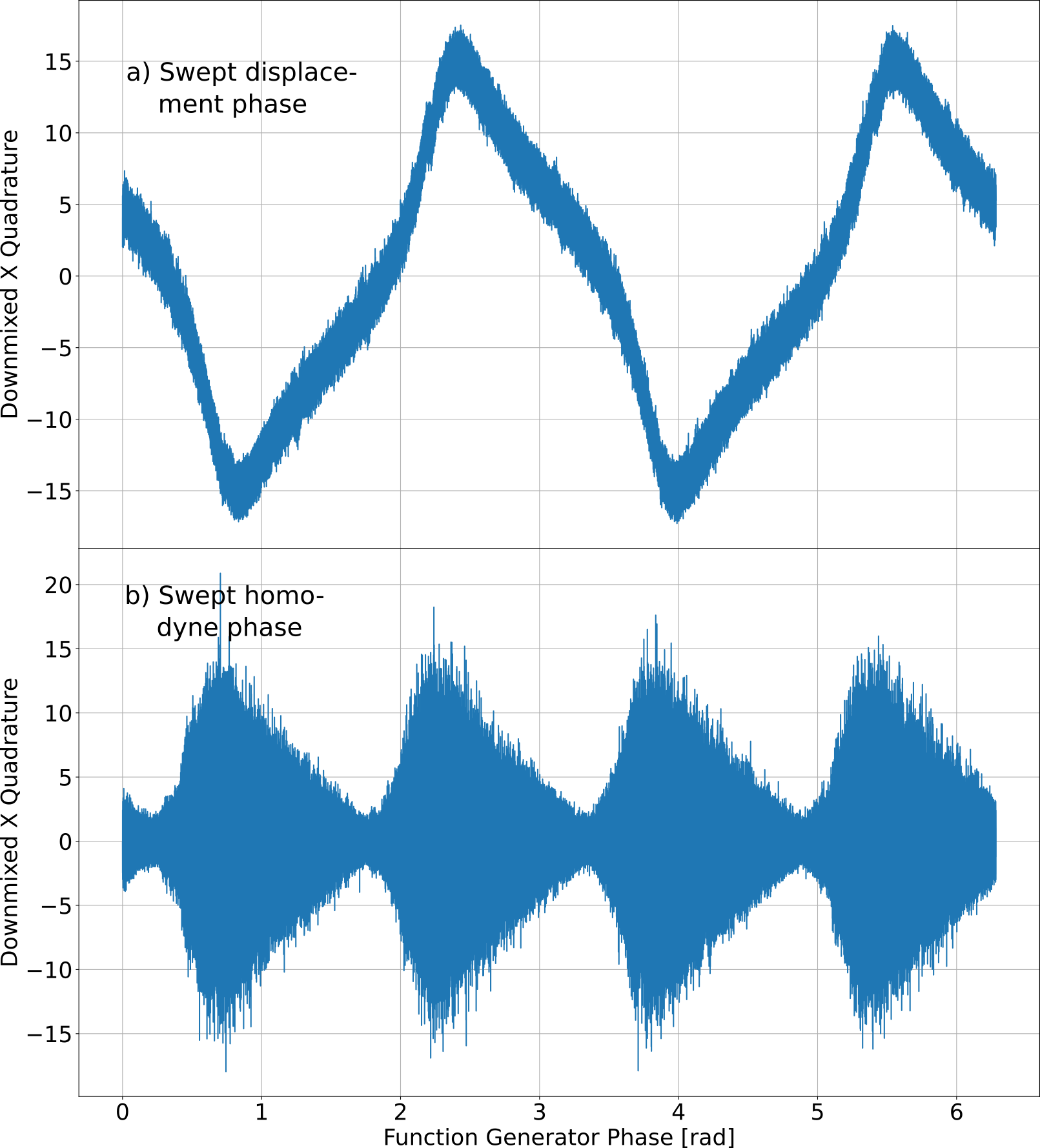}
    \caption{X quadrature measurements normalized to shot noise as a function of the swept ELO phase for a) the displacement phase lock and b) the homodyne phase lock. }
    \label{fig: angle_cali_raw}
\end{figure}

\subsection{Displacement and homodyne angles}
In the coherent locking scheme that stabilizes the phase between the squeezed light and the displacement (and the squeezed light and the local oscillator), the changing of the phase of the \SI{40}{\mega\hertz} electrical local oscillator that downmixes the error signal of the feedback loop rotates the angle $\phi_\alpha$ ($\phi_{HD}$). The correspondence between the set angle of the ELO and the actual quadrature angle is not linear, and thus needs to be calibrated. In order to do this, the phase of the displacement (homodyne) ELO function generator is swept through $2\pi$ over 2 seconds, while the homodyne output is recorded at \SI{50}{\mega\sample\per\second} and downmixed to the \SI{5}{\mega\hertz} sideband using a \SI{1}{\mega\hertz} lowpass filter. The data is normalized to the shot noise standard deviation. For the calibration of the displacement angle, a \SI{5}{\mega\hertz} displacement is added and the homodyne angle is manually set to the squeezed quadrature. For the homodyne angle calibration, the displacement is removed leaving only vacuum squeezing. The raw data of these measurements are shown in \cref{fig: angle_cali_raw}.

\begin{figure}
    \centering
    \includegraphics[width = 0.6\linewidth]{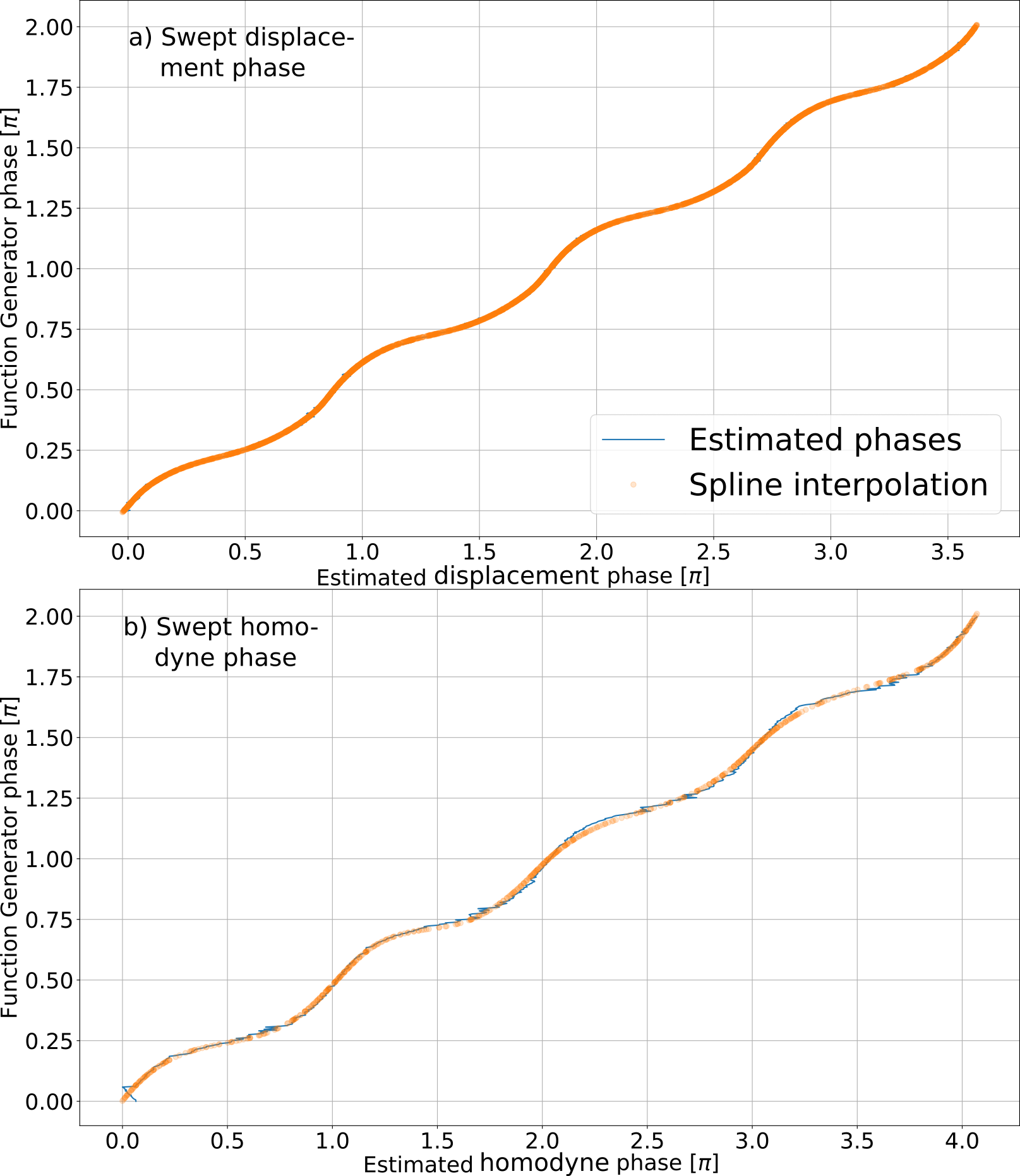}
    \caption{Plot of the function generator phase for a) the displacement phase lock and b) homodyne phase lock as a function of the estimated phases. The orange dots are the corresponding B-spline representations.}
    \label{fig: angle_cali_est}
\end{figure}

Using the data we can estimate the quadrature phase. For the measurement of the homodyne phase using squeezed vacuum, the method used in  \cite{Nielsen2023} can be directly applied, however for the displacement phase measurement, the marginal probability distribution has to be modified as
\begin{align}
P(X|\phi_\alpha) = \frac{1}{\sqrt{2\pi V_{-}}}e^{-\frac{\qty(X-2\abs{\alpha}\cos(\phi_\alpha))^2}{2V_{-}}}.
\end{align}
The estimated phases are unwrapped, and the function generator phases of the displacement (homodyne) ELOs are interpolated as a function of the estimated phases using a 3-order B-spline with a smoothing parameter of 1.3. The estimated phases and corresponding spline-representations are shown in \cref{fig: angle_cali_est}.

\begin{figure}
    \centering
    \includegraphics[width = 0.6\linewidth]{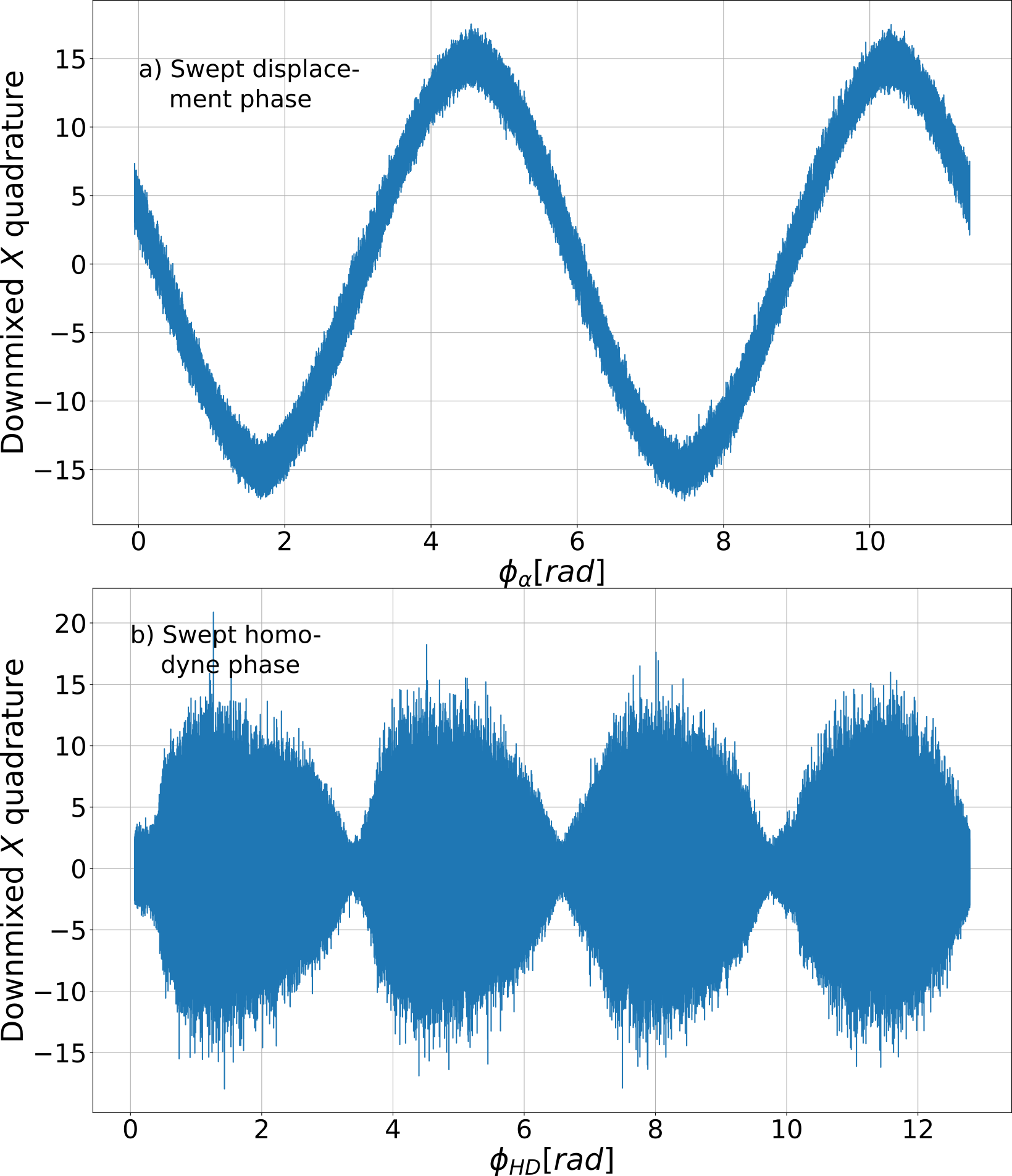}
    \caption{X quadrature measurements normalized to shot noise as a function of the estimated quadrature phase for a) the displacement phase lock and b) the homodyne phase lock.}
    \label{fig: angle_cali_corr}
\end{figure}

Using these B-spline interpolations, we can calibrate the system angles. Although the oscillatory behaviour of these calibrations generally result in the set angles to correspond to the desired angles, since the squeezing angle by itself is ill-defined, this can lead to the failure of these calibrations. This can result in a slight error between the set angle and the desired angle.

These calibrations can be verified by plotting the data in \cref{fig: angle_cali_raw} as a function of the unwrapped estimated phases instead of the function generator phase as shown in \cref{fig: angle_cali_corr}.

\section{Simulation model of the experiment}
The experimental setup was modelled analytically using Mathematica and the numerical simulation of the gradient descent-based variational quantum algorithm was done using the PennyLane \cite{bergholm2022pennylane} Python library. Considering it’s a continuous-variable system, a Gaussian quantum simulator was utilized as the backend.

A single-mode displaced squeezed state is prepared, characterized by a fixed degree of squeezing and displacement magnitude. The displacement angle, $\phi_\alpha$, and the homodyne detection angle, $\phi_{HD}$, are taken as the free parameters. The outputs from the quantum system are the mean and variance of the probed quadrature.\\
The initial probe state is thus,
\begin{equation}
\hat{\rho_0}(\phi_\alpha)=\hat{D}(\alpha,\phi_{\alpha}) \hat{S}(r,\phi_{r}) |0\rangle
\end{equation}
where $\hat{D}$ and $\hat{S}$ represent the displacement and quadrature squeezing operators respectively. $r$ is the degree of squeezing and $\alpha$ the magnitude of displacement, corresponding to the experiment.

\begin{figure}[ht]
    \centering
    \includegraphics[width = 0.8\linewidth]{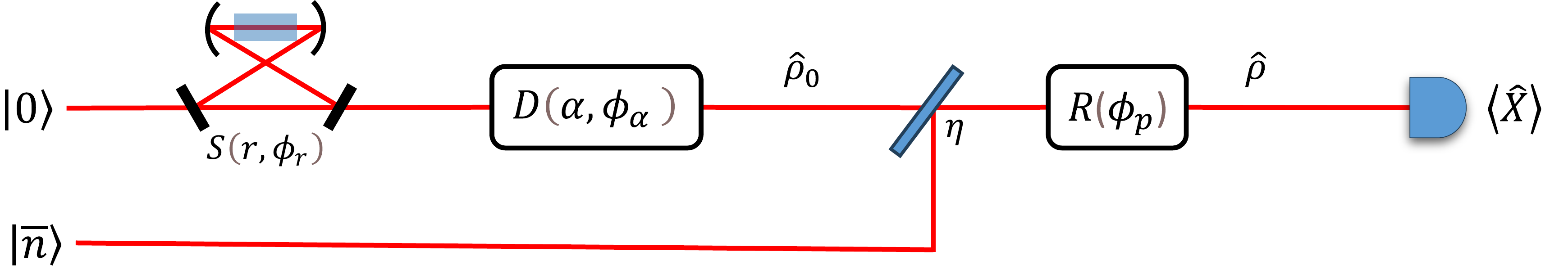}
    \caption{Schematic outline of the simulated system. Corresponding to the efficiency of the experimental system, the beamsplitter was simulated with a transmittivity of $\eta=0.72$ and no thermal photons ($\bar{n}=0$).}
    \label{fig: sim_model}
\end{figure}

To simulate the environmental interaction on the probe state, the photon loss and thermal noise in the system are modelled by coupling the probe state mode with a thermal noise mode $| \bar{n}\rangle$ in a fictitious beamsplitter with a transmittivity of $\eta$ (fig. \ref{fig: sim_model}). 
\begin{equation} 
\hat{\rho}=\sqrt{\eta} {\rho_0}+\sqrt{1-\eta} \; | \bar{n}\rangle
\end{equation}

\begin{figure}
    \centering
    \includegraphics[width = 0.8\linewidth]{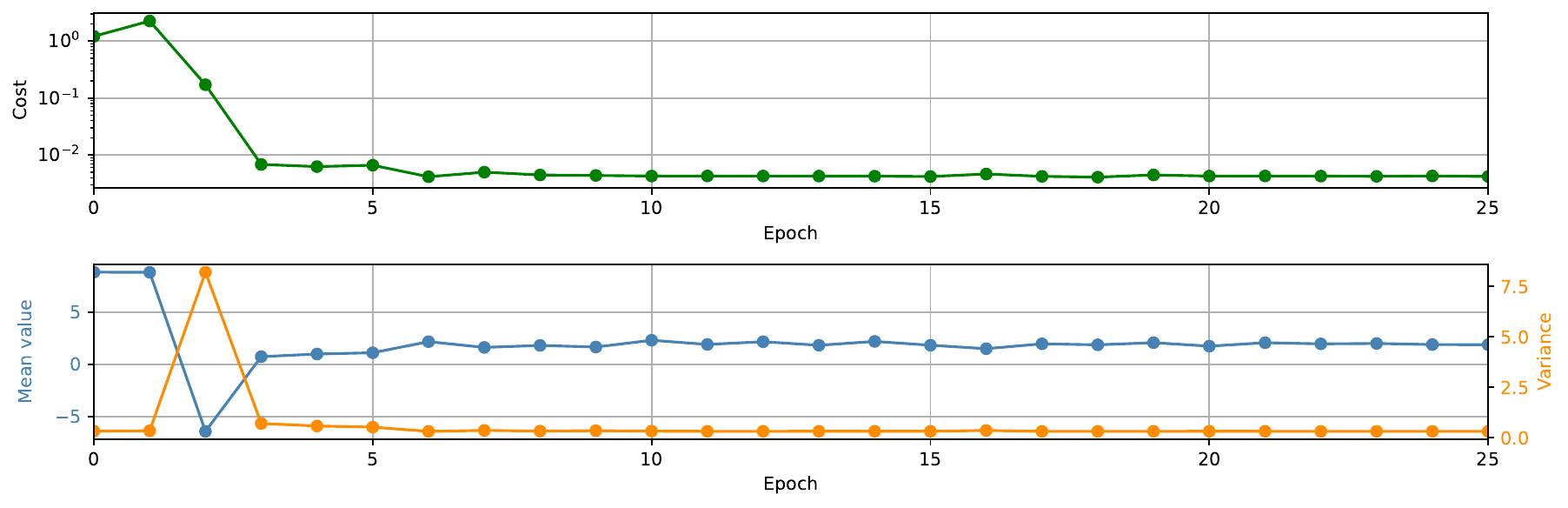}
    \caption{Simulation of the gradient descent-based variational algorithm on the experimental model with noise and loss. The experimental parameters are taken to be $r=1.52$, $|\alpha|=5.2$, and phase noise RMS = \SI{0.03}{\radian}. The cost function represents the single-shot Fisher Information and does not take into account the number of samples used, hence the higher value compared to the experiment.}
    \label{fig: sim_optimization}
\end{figure}

The phase noise in the measurement is modelled by encoding a random phase $\phi_{p}$, sampled from a Gaussian distribution centred around the root mean square value corresponding to the experimental scheme. The variational algorithm is then simulated by using a gradient descent optimization scheme to minimize the cost function as shown in fig. \ref{fig: sim_optimization}.
It is to be noted that choosing a suitable learning rate and initial parameters is necessary for the optimizer to converge to the minima without requiring a large number of epochs.

\begin{figure}[ht]
    \centering
    \includegraphics[width = 0.9\linewidth]{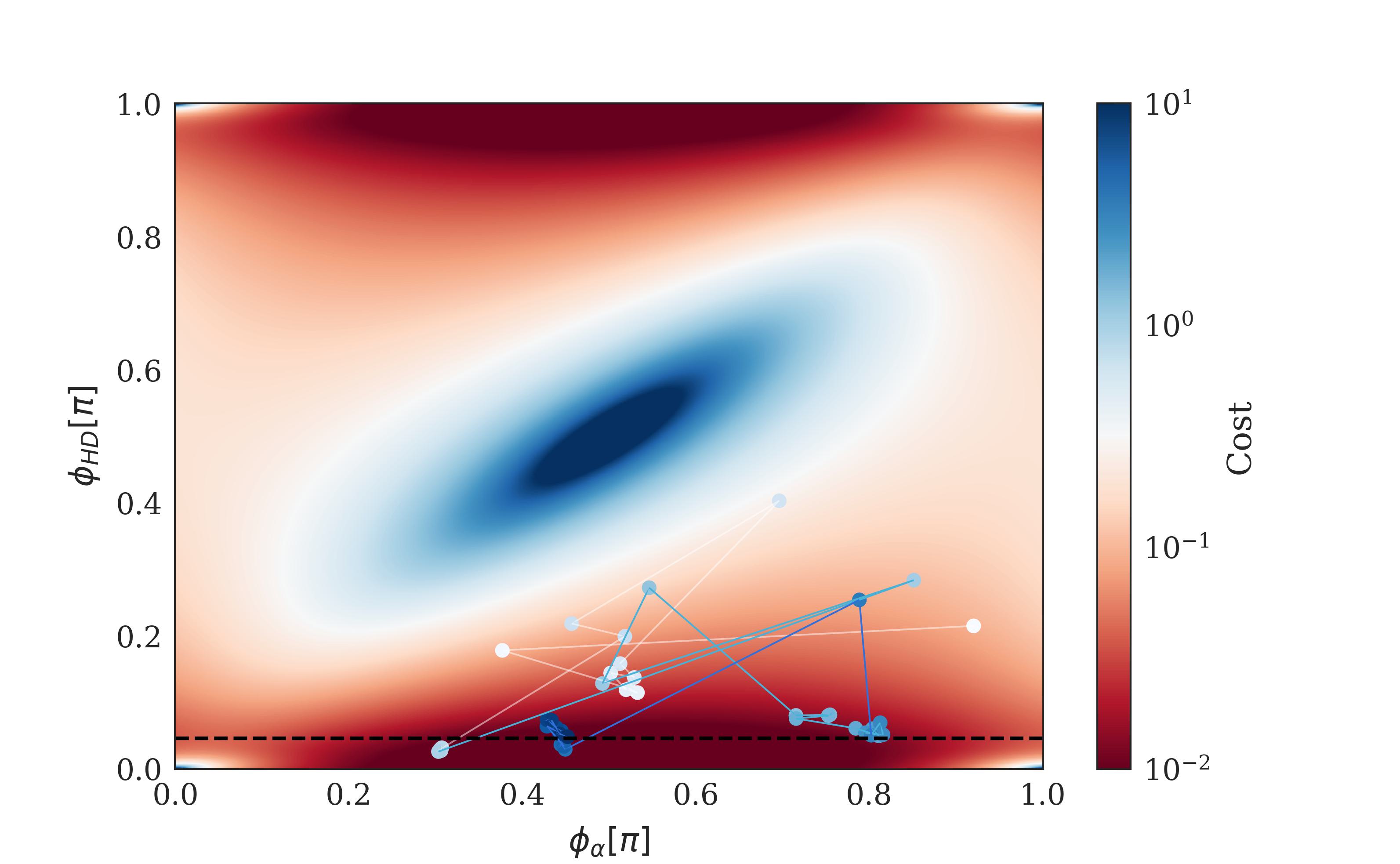}
    \caption{Simulated cost function landscape with kick measurement super-imposed. The points represent the cost for each epoch, with the colour intensity increasing progressively (the first epoch is depicted in white and the last epoch is the darkest blue). The black dotted line represents the theoretical optimum for the measurement angle $\phi_{HD}$}
    \label{fig:landscape_w_data}
\end{figure}

In \cref{fig:landscape_w_data} the cost function landscape is visualized by setting the simulation model with the experimentally determined parameters and probing the entire parameter space. The experimental data points from the kick-test described in the main text are superimposed on the landscape. We can observe that the optimizer effectively converges to the theoretical minimum.

\section{Bayesian optimization}
Bayesian optimization is an iterative and gradient-free way of estimating the global minimum $\xb_*$ of some function $f(\xb)$,

\begin{equation}
    \xb_* = \argmin_\xb f(\xb)
\end{equation}

In our experiment throughout the main paper, $\xb = [\phi_\alpha, \phi_{HD}]$ and $f(\xb)$ is the corresponding cost function value measured for those parameters. However, in many real-world experiments, $f(\xb)$ might be corrupted with additive noise, such that the only measurable quantity $y(\xb)$ is given by 

\begin{equation}
    y(\xb) = f(\xb) + \epsilon
\end{equation}
where it is often assumed that $\epsilon \sim \mathcal{N}(0,\sigma_{noise}^2)$. For low-dimensional $\xb$ (in our case only 2 dimensions) and expensive (e.g. time or monetary) queries to $y(\xb)$, Bayesian optimizer is a very relevant optimizer candidate \cite{shahriari2015taking}. This is indeed the case for our experiment outlined in the main text. In order to estimate $\xb_*$, the Bayesian optimizer framework needs two quantities: 1) a surrogate function and 2) an acquisition function. The surrogate function "mimics" the observed datapoints $y(\xb)$, and is thus referred to as a \textit{surrogate} function. Using this function, the acquisition function chooses which points to choose next by means of taking the maximum argument to the acquisition function. In our experiments, we choose a Gaussian Process as surrogate, and Expected Improvement as acquisition. In the next sections, we introduce these two quantities.

\subsection{Gaussian Process}
A Gaussian Process is a non-parametric model, which probabilistically models a variable $p(y_*|\xb_*)$ with a normal distribution. It does so by conditioning on the corresponding input pair $\xb_*$ as well as previously seen datapoints: a collection of $N$ input/output datapoints $(\xb_n, y_n)_{n=1}^N$ where $\xb_n$ is the $n$'th input and $y_n$ is the corresponding output. The entire collection of input datapoints, can be collected in an input matrix $\Xb$ and the corresponding outputs in a vector $\yb$. Specifically, the distribution over any output $y_*$, which together with the corresponding input $\xb_*$ we refer to as a testpoint, is given by the normal distribution 

\begin{equation}
    p(y_*|\xb_*,\Xb,\yb) = \mathcal{N}(\mub_{y_*|\xb_*,\Xb,\yb},  \Sigmab_{y_*|\xb_*,\Xb,\yb}).
        \label{eq:gp-pred-dist}
\end{equation}

where 
\begin{align}
            \mub_{y_*|\xb_*,\Xb,\yb} &= K(\xb_*,\Xb)^\top[K(\Xb,\Xb) + \sigma_n^2 I]^{-1} \yb, \\
            \Sigmab_{y_*|\xb_*,\Xb,\yb} &= K(\xb_*,\xb_*) - K(\xb_*,\Xb)^\top [K(\Xb,\Xb) + \sigma_n^2 I]^{-1} K(\Xb,\xb_*).
        \label{eq:mean-uncertainty-gp}
\end{align}
and where $K(\Xb,\Xb)$ is called the kernel matrix. The kernel matrix is an $N \times N$ positive semi-definite matrix that contains pairwise similarity measures between the training points (vectors). Similarly $K(\xb_*,\Xb)$ is a $N$ dimensional vector with a similarity measure between the test point and all the training points. It is easily verified from \cref{eq:mean-uncertainty-gp} that both $\mub_{y_*|\xb_*,\Xb,\yb}$ and $\Sigmab_{y_*|\xb_*,\Xb,\yb}$ are scalars. For full derivation, we refer to \cite{foldager2023quantum}.

A popular choice of similarity measure between two datapoints (represented as vectors) is the Radial Basis Function (RBF) Kernel (also called a Gaussian Kernel) given by
\begin{equation}
    K_{RBF}(\xb,\xb') = exp\left(-\frac{||\xb-\xb'||^2}{2 \sigma^2}\right)
\end{equation}
which is a function where similarity exponentially decays as the Euclidean distance between two points increases. The hyperparameter $\sigma$ is called the length scale, as it defines the scale of how quickly the similarity should decrease. In our experiments, we use a Logarithmic Normal distribution prior and estimate it by maximizing the log-likelihood of the Gaussian Process. We also include an output scale hyperparameter $\sigma_{scale}$ such that the final kernel function is given by $\sigma_{scale}\cdot K_{RBF}(\xb,\xb')$.

\subsection{Acquisition function}
We use 136 datapoints and refer to this as our initial training set $\mathcal{D}_0 := [\Xb, \yb]$ and using the above equations we get the predictive distribution of the Gaussian Process. We refer to this Gaussian Process as our surrogate model, since it models the underlying loss landscape. 

We now iteratively query points by finding the input point $\xb_*$ that maximizes an acquisition function. A popular choice for acquisition function to go together with the Gaussian Process surrogate model is the expected improvement given by

\begin{equation}
            \text{EI}(\xb_*) =
            (\mu(\xb_*) - f(\xb_*^+))\Phi(Z(\xb_*))+\sigma(\xb_*)\phi(Z(\xb_*)), \label{eq:ei}
\end{equation}

where $f(\xb_*^+)$ is the current best guess of the global minimum and $\xb_*^+$ is the parameter setting, $\Phi$ is the cumulative distribution function of a standard normal distribution, $\phi$ is the probability density function of a standard normal distribution and $\mu(\xb_*)$ and $\sigma(\xb_*)$ comes from the surrogate predictive distribution. EI is based on calculating expected marginal gain utility in the Gaussian Process after performing observation for candidate parameters \cite{bayesian2023}. The next queried input is thus given by

\begin{equation}
    \xb_{next} = \argmax_{\xb} \text{EI}(\xb)
\end{equation}

Note that $\xb_{next}$ is a parameter combination $[\phi_{HD}, \phi_{\alpha}]$ which we have not used before. This parameter set is now used in the experiment to get the corresponding loss value $y(\xb_{next})$. The dataset is now updated with this value to obtain $\mathcal{D}_1$, and so on. If the reader wants a pedagogical illustration of this process, have a look at Fig. 2.6 in \cite{foldager2023quantum}.

\begin{figure*}
    \centering\includegraphics{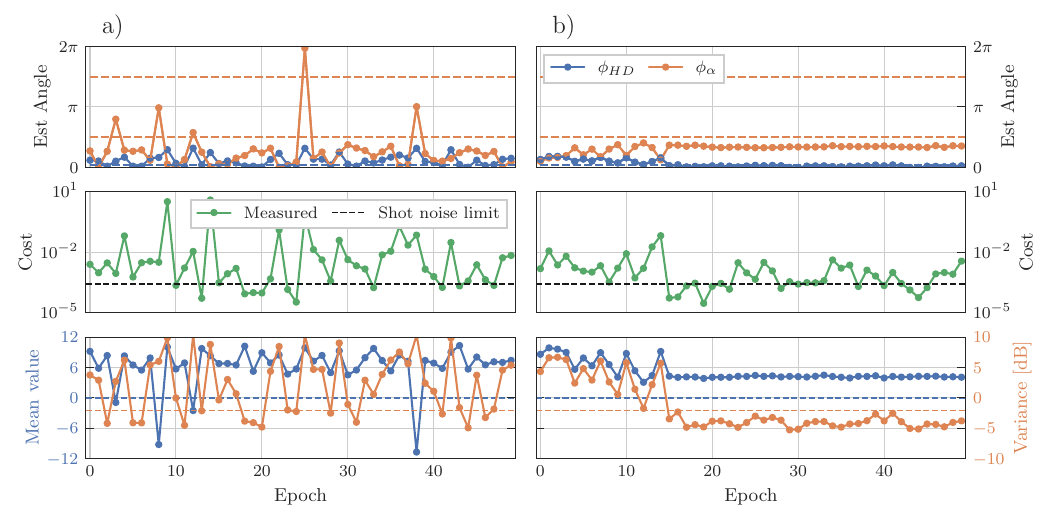}
    \caption{Demonstration of Bayesian optimization over 50 epochs for a) very high and b) very low mean and variance of the hyper-parameters of the Gaussian Process. (Top) The phase angles set by the algorithm. (Middle) The measured cost function $C = 1/F$. The dotted line is the shot noise limit taking into account the number of photons in the measurement and the number of samples used to estimate the cost function. (Bottom) The measured quadrature mean values and variances. The dotted lines are the optimal values predicted by the theory (Supplement section I).}
    \label{fig:NoGrad_supp}
\end{figure*}

\subsection{Hyper-parameters of Gaussian Process}
Hyperparameters of the Gaussian Process were chosen very carefully. Their wrong definition might lead to the model's worse performance or in a critical case, the model not being able to perform optimization at all. Defining hyperparameters involves setting values of the mean and standard deviation of the aforementioned length scale and output scale hyperparameters. They can be either well-defined values (as for presented results in the main section of the paper), too strict (too low) or too loose (too high). On \cref{fig:NoGrad_supp} a) we can see results for too loose definition of hyper-parameters and on  \cref{fig:NoGrad_supp} b) results for too strict values defining hyper-parameters are presented.

In figure \cref{fig:NoGrad_supp} b) we see that too strict values defining hyperparameters lead to the optimizer performing very small changes of angle parameters resulting in it being stuck in areas that not necessarily are optimum. This argument is supported by relatively high values of a cost function which rarely goes below the limit of shot noise. In figure \cref{fig:NoGrad_supp} a) we also see that too loose values of mean and variance of hyperparameters result in significant changes of angles in consecutive epochs. It is followed by considerable changes in the mean value of X quadrature, its variance relative to the shot noise and in values of a cost function as well. This behaviour is an example of a completely wrong choice of hyperparameters for which the model can't relate its definition with observations gathered during measurements. It results in the model "randomly walking" over the domain of angle parameters and not being able to perform optimization.

Both of these cases show the importance of defining the model correctly and prove that it should be done with meticulous attention.

\bibliographystyle{ieeetr}
\bibliography{Supp}